\documentclass{article} 
\usepackage{iclr2023_conference,times}
\usepackage{hyperref}
\usepackage{url}
\usepackage{graphicx}
\usepackage{subfig}
\usepackage{wrapfig}
\usepackage{mathabx}

\title{Sea level Projections with Machine Learning using Altimetry and Climate Model ensembles}

\author{%
Saumya Sinha\thanks{Corresponding author: Saumya Sinha,  saumya.sinha@colorado.edu} \\
  University of Colorado\\
  Boulder, CO, USA \\
  \And
  John Fasullo  \\
  National Center for Atmospheric Research\\ 
  Boulder, CO, USA \\
   \And
 R. Steven Nerem \\
 University of Colorado\\
  Boulder, CO, USA \\
  \And
 Claire Monteleoni \\
 University of Colorado\\
  Boulder, CO, USA \\
}

\iclrfinalcopy 
\begin{document}

\maketitle

\begin{abstract}
Satellite altimeter observations retrieved since 1993 show that the global mean sea level is rising at an unprecedented rate (3.4mm/year). With almost three decades of observations, we can now investigate the contributions of anthropogenic climate-change signals such as greenhouse gases, aerosols, and biomass burning in this rising sea level. We use machine learning (ML) to investigate future patterns of sea level change. To understand the extent of contributions from the climate-change signals, and to help in forecasting sea level change in the future, we turn to climate model simulations. This work presents a machine learning framework that exploits both satellite observations and climate model simulations to generate sea level rise projections at a 2-degree resolution spatial grid, 30 years into the future. We train fully connected neural networks (FCNNs) to predict altimeter values through a non-linear fusion of the climate model hindcasts (for 1993-2019). The learned FCNNs are then applied to future climate model projections to predict future sea level patterns. We propose segmenting our spatial dataset into meaningful clusters and show that clustering helps to improve predictions of our ML model.


\end{abstract}

\section{Introduction}
With melting ice sheets and the growing warmth of the ocean water, the global mean sea level is rising at an extraordinary rate and is accelerating (0.08mm/year$^2$) 
 [\citenum{nerem2018climate}]. While, on average, the sea level has risen 10 cm over the last 30 years, there is a considerable variation in the regional rates of the sea level change [\citenum{hamlington2016ongoing}]. These regional
patterns are richer in information and can be very useful in examining the impact of climate-driven factors including greenhouse gases, industrial aerosols, and biomass burning in the sea level change. Studies in [\citenum{fasullo2018altimeter,fasullo2020sea,fasullo2020forced}] have found the regional variations to be linked to
aerosols and greenhouse gas emissions. Identifying how regional trends will evolve in the future is also beneficial for socioeconomic planning. Our work investigates the sea level at a 2-degree or 180 x 90 (longitude x latitude) spatial resolution.  

Now that there are almost three decades of satellite altimeter records or observations\footnote{We obtain our altimeter data for the period 1993-2019 from \url{https://www.aviso.altimetry.fr/en/data/products/ocean-indicators-products/mean-sea-level.html}}, we want to explore the rising rate of the sea level further and investigate how much of this rise can be attributed to climate-change signals. To do
this, we turn to the climate models in order to provide a better understanding of the altimeter data and to help in forecasting future sea level change by learning more about the extent of the causal contributions from such factors. A climate model uses mathematical equations to simulate complex physical and chemical processes of Earth systems such as the atmosphere, land, ocean, ice, and solar energy [\citenum{ucarclmwebsite}]. A recent study by ~\citet{fasullo2018altimeter} with two climate model large ensembles showed that the forced responses of greenhouse gas and aerosols have begun to emerge in the altimeter data patterns. This motivates us to include climate models in designing our machine learning (ML) pipeline.

 Our goal is to generate sea level projections 30 years into the future and at a 2-degree spatial resolution utilizing the altimeter data as well as climate model simulations.  
 Some past works have used satellite altimeter data and adopted ML techniques to perform sea level prediction. While tide-gauge data has also been used for similar tasks, satellite altimetry provides nearly global coverage.~\citet{braakmann2017sea} used a combination of CNN + ConvLSTM layers to perform interannual sea level anomalies (SLA) prediction 
 over the Pacific Ocean. [\citenum{sun2020estimation}] work with LSTM for South China Sea. In [\citenum{balogun2021sea}], the authors include ocean-atmospheric features like sea surface temperature, salinity, surface atmospheric pressure to
build support vector and LSTM models for the West Peninsular Malaysia coastline. [\citenum{nieves2021predicting}] make use of gaussian processes and
LSTM to predict sea level variation along the regional coastal zones. In [\citenum{hassan2021comparative}], they compare various machine learning techniques to predict global mean sea level rise. However, none of these models go so far as to forecast sea level change \emph{30} years in advance. They also do not produce forecasts for the entire globe. Our work addresses the problem at a much bigger spatial scale that includes all the oceans and a much longer time horizon in the future. Our framework utilizes climate models to understand the causal effects of climate-change signals. We add interpretability to our model by explaining the contributions of the climate model ensembles. We also present a way to segment spatial data and show the performance of our method for predicting the sea rise trend. We do a quantitative evaluation of the predictions on the past data and introduce a qualitative analysis of the predictions on the future data while also producing uncertainty estimates associated with the ML prediction.

 {\bf Pathways to Climate Impact} Forecasting long-term sea level change is a complex problem given the natural variability of the ocean, the wide range of processes involved, and complex non-linear interactions playing a role in the sea level change. Through our work, we show the potential of machine learning in producing 
these forecasts 30 years in advance and over all oceans at a reasonably good spatial resolution. This is a promising step towards understanding the impact of climate change on the current sea level rise and how it can influence its future course. The longer horizon prediction in the future can be vital in providing us with more time to plan and adapt to the rising sea level.

\section{Dataset and Problem formulation}
\begin{figure}[h]
\begin{center}
\includegraphics[scale=0.35]{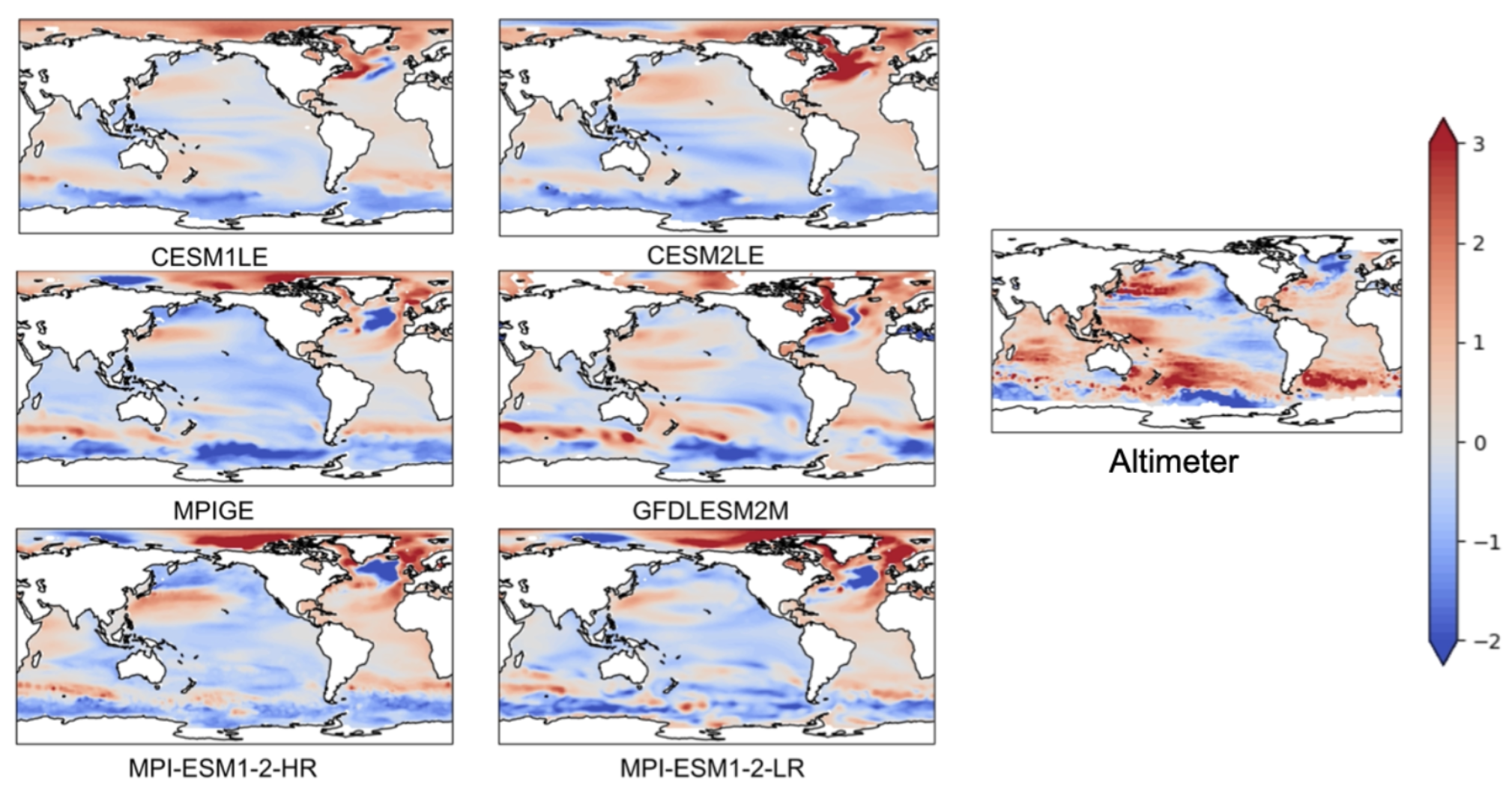}
\end{center}
\vspace{-3mm}
\caption{\small Figure shows the sea level trend maps for the 6 climate model ensembles (on left) and satellite altimeter data (on right) for the period 1993-2019. Here the trend values are visualized in mm/year.}
\label{label:trend_climate_models_and_altimeter}
\end{figure}
Our altimeter dataset is a monthly sea surface height (SSH) data at 1/4-degree spatial resolution for the time period 1993-2019. For the same duration, we obtain monthly SSH at 1-degree spatial resolution from 6 different climate model ensembles. They are: CESM1 (large ensemble) [\citenum{kay2015community}], CESM2 (large ensemble) [\citenum{danabasoglu2020community}], MPIGE [\citenum{maher2019max}], GFDLESM2M [\citenum{dunne2013gfdl}], MPI-ESM1-2-HR [\citenum{muller2018higher}] and MPI-ESM1-2-LR [\citenum{giorgetta2013climate}]. 
Model simulations for individual members of the model ensembles are averaged to create the sea surface height (SSH) variable we work with. The spatial SSH fields for both the altimeter data and climate models are regridded to a 2 degree, i.e a 180x90 grid (as it speeds up the computation while still keeping a reasonable resolution), and their global mean is removed as well. For every ocean grid point, a linear trend is fitted to the monthly SSH time series for the 1993-2019 time period. This way, we obtain a \textit{single} trend map, for all the climate model ensembles and the altimeter (see Figure~\ref{label:trend_climate_models_and_altimeter}). Working with trends helps to avoid the monthly variability of the SSH fields. This can be beneficial to our machine learning model as removing other variabilities can help it to learn from the climate-change signals better. 

We flattened the spatial grid to create our dataset. For every ocean lat/long there is a trend (in cm/year) provided by each of the 6 climate models, comprising the input (X) and also from the altimeter (ground truth), which is the label (Y) for our ML training. We frame our problem as a supervised machine learning task that utilizes both climate model hindcast trends, and altimeter trend while absorbing the biases that the climate models have away from the altimeter data. Figure~\ref{label:trend_climate_models_and_altimeter} shows that the climate models don't reproduce the trend pattern in altimeter data very well and they also disagree with each other substantially. There is also a difference in variability between the altimeter and the climate models as can be observed in Figure~\ref{label:trend_climate_models_and_altimeter} and we expect our method to address this.
It is worth noting that we do not have altimeter records for all the latitudes. They are roughly present for latitudes [$-70$ to +$70$] that give us $7,643$ data points excluding land grid points.

While we have access to long periods of climate model simulations in the past as well as the future, our learning setup as described in [\citenum{sinha2022sea}] using an UNet-based spatiotemporal forecasting model trained on the climate model dataset did not perform very well in forecasting on the altimeter data. This is expected as there is a lot more variability in the altimeter trend as compared to the climate models (Figure~\ref{label:trend_climate_models_and_altimeter}). Moreover, with just a single altimeter trend map (1993-2019), we are limited by the data in the temporal dimension. This motivated us to exploit a reasonably dense spatial data instead and formulate our problem as described above.

\section{Method}
\label{method}
Our supervised machine learning pipeline is trained for the period 1993-2019. We get the supervision from the altimeter trend and inputs or features to our ML model is provided by the 6 climate model hindcast trends for the same time period. 
In the inference phase, we predict trend projections for 30 years later. This is done by taking the climate model projected trends for 2023-2049 and passing them through the learned ML model to produce a projected altimeter trend. 
Our ML model is a fully connected neural network (FCNN) trained with mean squared error (MSE) as the loss. The MSE is weighted where the weights are the $abs(cosine)$ of the latitude of the grid points to give less weight to polar regions as compared to other regions. 

\begin{figure}[!htb]
\vspace{-4mm}
 \centering
 \subfloat[K-means]{%
      \includegraphics[width=0.29\textwidth]{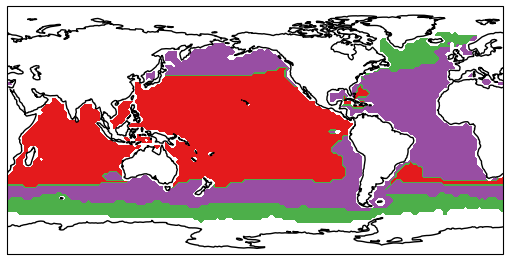}}
 \qquad
 \subfloat[Spectral]{%
      \includegraphics[width=0.29\textwidth]{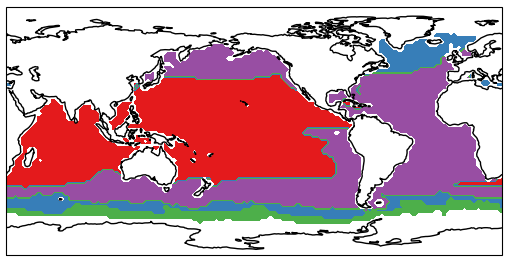}}
 \qquad
 \subfloat[Hard-coded]{%
      \includegraphics[width=0.29\textwidth]{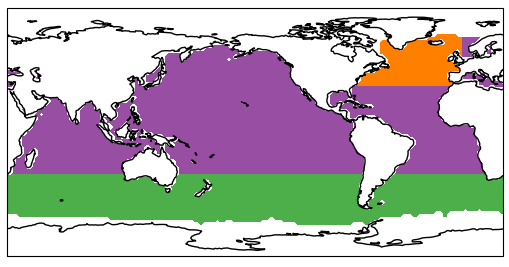}}
\vspace{-2mm}
\caption{\small Clusters obtained from (a) K-means clustering, (b) Spectral clustering, and (c) a hard-coded clustering derived from our physical understanding where the North Atlantic Ocean is assigned a single cluster, latitudes from south up to -30 are assigned another cluster and the rest belong to the 3rd cluster.}
\label{label:clusters}
\vspace{-2mm}
\end{figure}

\textbf{Clustering}:
We segment our spatial grid into clusters with various clustering methods and observe the performance of the ML model when trained conditioned on these clusters i.e a separate FCNN is trained for \emph{each} cluster. The time series of the altimeter sea-surface height serves as the features for K-means clustering and Spectral clustering. We also perform a clustering scheme that is derived from our physical knowledge of the data (described in Figure~\ref{label:clusters}). The spatial segmentations with all three clustering approaches can be seen in Figure~\ref{label:clusters}. These approaches are compared to each other and also with a setup where the spatial grid is not segmented at all. We make use of k-fold cross-validation (k=5) to choose the best hyperparameters for each cluster, ending up with different FCNN architectures per cluster. To elaborate further, each of the red and purple clusters in the spectral clustering setting as seen in Figure~\ref{label:clusters}(b), learns an FCNN consisting of 3 hidden layers with 1024, 512, and 256 neurons respectively. For each of the other two smaller clusters, we use an FCNN with 2 hidden layers and 256, 128 neurons respectively. 
 
\section{Results}
We report different metrics for past and future time periods.\\
\textbf{1) 1993-2019 (with ground truth available) : }
We show the RMSE scores for the three clustering approaches and a no clustering setup (Table~\ref{results-table}). Table~\ref{results-table} also shows the Pearson correlation scores between the ML predicted trend and the true altimeter trend for 1993-2019. Comparing their cross-validation scores (as described in \ref{method}) also gives a similar performance order. The RMSE and correlation scores are weighted as described in \ref{method}.
We observe that \textit{K-means Clustering} outperforms others. \textit{Kmeans} and \textit{Spectral} have similar scores, this may be due to the fact that they have almost similar clusters except for a very small segmented region in the Southern Ocean in the spectral clusters. 
Their scores are also better than the \textit{Hardcoded clustering} and \textit{No clustering}. 
We also examine each of the segmented regions by looking at each cluster's RMSE and correlation scores. 
\begin{wraptable}{r}{0.5\textwidth}
\caption{Table comparing the performance in terms of weighted RMSE (in mm/year) and correlation for the clustering methods for 1993-2019.}
\label{results-table}
\begin{center}
\vspace{-3mm}
\begin{tabular}{lll}
\multicolumn{1}{c}{\bf Method}  &\multicolumn{1}{c}{\bf RMSE$\downarrow$} &\multicolumn{1}{c}{\bf Correlation$\uparrow$}
\\ \hline \\
No Clustering          &0.57 & 0.9\\
Spectral Clustering             &0.48 & 0.93\\
\textbf{K-means Clustering}             &\textbf{0.44} & \textbf{0.93}\\
Hardcoded Clustering           &0.55 & 0.89\\
\end{tabular}
\end{center}
\vspace{-4mm}
\end{wraptable}
We derive more insights by visualizing the difference map between the true trend and prediction. The North Atlantic and Southern Ocean in general show higher differences. Some of the higher difference zones in the map could be caused by ocean eddies. Our ML model's performance is much better in the Pacific, which is significant as it is a crucial area for socioeconomic impacts. \\
\textbf{Interpretability:} In order to add interpretability to the models, we use SHAP [\citenum{lundberg2017unified}] values to explain the contributions of the climate model ensembles which are feature inputs to our FCNNs. For \textit{Kmeans} and \textit{Spectral clustering}, on the "red" cluster of Figure~\ref{label:clusters}(a,b) (which we are most interested in) SHAP values indicate the climate model CESM1 (large ensemble) to be the most important, followed by CESM2 (large ensemble) and MPIGE. 

\begin{figure}[!htb]
 \centering
 \subfloat[]{%
      \includegraphics[width=0.29\textwidth]{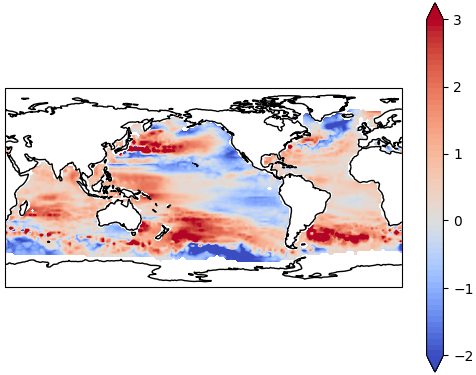}}     
 \qquad
 \subfloat[]{%
      \includegraphics[width=0.29\textwidth]{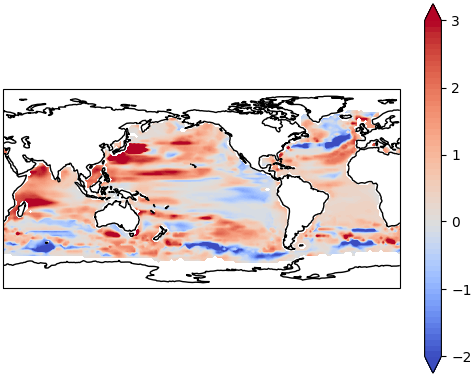}}
\qquad
 \subfloat[]{%
      \includegraphics[width=0.29\textwidth]{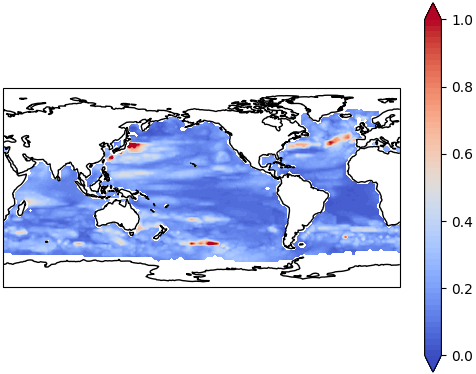}}
\vspace{-2mm}
\caption{\small The trend estimates are in mm/year. Figure (a) shows the altimeter trend in 1993-2019 with RMS 1.28mm/year, (b) shows the trend predicted with ML for the future 2023-2049 with RMS 1.24mm/year and (c) shows the ML model uncertainty in terms of the standard deviation (mm/year).}
\label{label:prediction_outcomes}
\end{figure}

\textbf{2) 2023-2049 : }
It is harder to gauge the performance of any ML method without the ground truth. In this case, we do a qualitative analysis of the predicted trend in terms of cumulative variability and we expect our ML models to predict trends with variability similar to the variability of the 1993-2019 altimeter trend. We take the root mean square (RMS) of the trend to quantify the notion of variability in the trend. Figure~\ref{label:prediction_outcomes}(a) shows that the altimeter trend from 1993-2019 (which is referred to as \textit{persistence} in climate literature) has a very high variability \textbf{(1.28mm/year)}.
In figure~\ref{label:prediction_outcomes}(b), we show the future trend predictions obtained from the \textit{K-means clustering} model and inspect its variability. Figure~\ref{label:prediction_outcomes}(b) shows a high variability \textbf{(1.24mm/year)} in our predicted trend for 2023-2049, though it is still lesser than the altimeter trend variability of the past.
If we look at the predicted variability concerning other clustering methods, we found that all of them including \textit{Hardcoded clustering} predictions showed a higher variability (1.21mm/year) compared to the \textit{No clustering} setting (0.99mm/year). This analysis strengthens our hypothesis to segment the spatial grid and learn an ML model conditioned on each segmented region for more optimal predictions that can capture better variability. 
\\
\textbf{Model uncertainty:} Additionally, our framework provides uncertainty over the predictions given by the ML model. Our FCNN model includes dropout layers to reduce overfitting while training. We use the Monte Carlo dropout [\citenum{gal2016dropout}] approach where in the inference phase, we perform multiple forward passes (with different dropout masks) through our ML model and report the mean of the ensemble of predictions as the prediction outcome and their standard deviation as the model uncertainty (this is shown in Figure~\ref{label:prediction_outcomes}(c)).
\section{Discussion}
 This work showcases the efficacy of machine learning for the crucial task of long-term sea level prediction on a 2-degree spatial grid leveraging the forecasts provided by climate model ensembles. Fully connected neural networks learn to map climate model projections to future trends and their superior performance, especially in the Pacific Ocean shows promise in this application. K-means clustering is found to have the lowest training RMSE, but the predictions generated for the future have lower variability as compared to the altimeter persistence. We need to investigate this further as we expect to see an increase in trend variability in the future that aligns with the climate model projections for 2023-2049 that show an increased variability as compared to their 30 years' past values. We plan to use spatial smoothening\footnote{\url{https://www.ncl.ucar.edu/Document/Functions/Built-in/exp_tapersh.shtml}} to our altimeter data to counter the influence of small-scale ocean eddies and potentially produce predictions with higher variability. The domain experts in our team will also analyze the ML model predictions and uncertainty further and compare it across all the clustering approaches. Our initial results look promising, furthering the need for more research on anthropogenic climate change and how it can impact future sea level rise.






\setcitestyle{numbers}
\bibliographystyle{plainnat}
\bibliography{main}

\begin{thebibliography}{20}
\providecommand{\natexlab}[1]{#1}
\providecommand{\url}[1]{\texttt{#1}}
\expandafter\ifx\csname urlstyle\endcsname\relax
  \providecommand{\doi}[1]{doi: #1}\else
  \providecommand{\doi}{doi: \begingroup \urlstyle{rm}\Url}\fi

\bibitem[Balogun and Adebisi(2021)]{balogun2021sea}
Abdul-Lateef Balogun and Naheem Adebisi.
\newblock Sea level prediction using arima, svr and lstm neural network:
  assessing the impact of ensemble ocean-atmospheric processes on models’
  accuracy.
\newblock \emph{Geomatics, Natural Hazards and Risk}, 12\penalty0 (1):\penalty0
  653--674, 2021.

\bibitem[Braakmann-Folgmann et~al.(2017)Braakmann-Folgmann, Roscher, Wenzel,
  Uebbing, and Kusche]{braakmann2017sea}
Anne Braakmann-Folgmann, Ribana Roscher, Susanne Wenzel, Bernd Uebbing, and
  J{\"u}rgen Kusche.
\newblock Sea level anomaly prediction using recurrent neural networks.
\newblock \emph{arXiv preprint arXiv:1710.07099}, 2017.

\bibitem[Danabasoglu et~al.(2020)Danabasoglu, Lamarque, Bacmeister, Bailey,
  DuVivier, Edwards, Emmons, Fasullo, Garcia, Gettelman,
  et~al.]{danabasoglu2020community}
Gokhan Danabasoglu, J-F Lamarque, J~Bacmeister, DA~Bailey, AK~DuVivier, Jim
  Edwards, LK~Emmons, John Fasullo, R~Garcia, Andrew Gettelman, et~al.
\newblock The community earth system model version 2 (cesm2).
\newblock \emph{Journal of Advances in Modeling Earth Systems}, 12\penalty0
  (2):\penalty0 e2019MS001916, 2020.

\bibitem[Dunne et~al.(2013)Dunne, John, Shevliakova, Stouffer, Krasting,
  Malyshev, Milly, Sentman, Adcroft, Cooke, et~al.]{dunne2013gfdl}
John~P Dunne, Jasmin~G John, Elena Shevliakova, Ronald~J Stouffer, John~P
  Krasting, Sergey~L Malyshev, PCD Milly, Lori~T Sentman, Alistair~J Adcroft,
  William Cooke, et~al.
\newblock Gfdl’s esm2 global coupled climate--carbon earth system models.
  part ii: carbon system formulation and baseline simulation characteristics.
\newblock \emph{Journal of Climate}, 26\penalty0 (7):\penalty0 2247--2267,
  2013.

\bibitem[Fasullo and Nerem(2018)]{fasullo2018altimeter}
John~T Fasullo and R~Steven Nerem.
\newblock Altimeter-era emergence of the patterns of forced sea-level rise in
  climate models and implications for the future.
\newblock \emph{Proceedings of the National Academy of Sciences}, 115\penalty0
  (51):\penalty0 12944--12949, 2018.

\bibitem[Fasullo et~al.(2020{\natexlab{a}})Fasullo, Gent, and
  Nerem]{fasullo2020sea}
John~T Fasullo, Peter~R Gent, and R~Steven Nerem.
\newblock Sea level rise in the cesm large ensemble: The role of individual
  climate forcings and consequences for the coming decades.
\newblock \emph{Journal of Climate}, 33\penalty0 (16):\penalty0 6911--6927,
  2020{\natexlab{a}}.

\bibitem[Fasullo et~al.(2020{\natexlab{b}})Fasullo, Gent, and
  Nerem]{fasullo2020forced}
John~T Fasullo, Peter~R Gent, and RS~Nerem.
\newblock Forced patterns of sea level rise in the community earth system model
  large ensemble from 1920 to 2100.
\newblock \emph{Journal of Geophysical Research: Oceans}, 125\penalty0
  (6):\penalty0 e2019JC016030, 2020{\natexlab{b}}.

\bibitem[for Science~Education()]{ucarclmwebsite}
UCAR~Center for Science~Education.
\newblock Climate modeling.
\newblock
  \url{https://scied.ucar.edu/learning-zone/how-climate-works/climate-modeling}.

\bibitem[Gal and Ghahramani(2016)]{gal2016dropout}
Yarin Gal and Zoubin Ghahramani.
\newblock Dropout as a bayesian approximation: Representing model uncertainty
  in deep learning.
\newblock In \emph{international conference on machine learning}, pages
  1050--1059. PMLR, 2016.

\bibitem[Giorgetta et~al.(2013)Giorgetta, Jungclaus, Reick, Legutke, Bader,
  B{\"o}ttinger, Brovkin, Crueger, Esch, Fieg, et~al.]{giorgetta2013climate}
Marco~A Giorgetta, Johann Jungclaus, Christian~H Reick, Stephanie Legutke,
  J{\"u}rgen Bader, Michael B{\"o}ttinger, Victor Brovkin, Traute Crueger,
  Monika Esch, Kerstin Fieg, et~al.
\newblock Climate and carbon cycle changes from 1850 to 2100 in mpi-esm
  simulations for the coupled model intercomparison project phase 5.
\newblock \emph{Journal of Advances in Modeling Earth Systems}, 5\penalty0
  (3):\penalty0 572--597, 2013.

\bibitem[Hamlington et~al.(2016)Hamlington, Cheon, Thompson, Merrifield, Nerem,
  Leben, and Kim]{hamlington2016ongoing}
BD~Hamlington, SH~Cheon, PR~Thompson, MA~Merrifield, RS~Nerem, RR~Leben, and
  K-Y Kim.
\newblock An ongoing shift in pacific ocean sea level.
\newblock \emph{Journal of Geophysical Research: Oceans}, 121\penalty0
  (7):\penalty0 5084--5097, 2016.

\bibitem[Hassan et~al.(2021)Hassan, Haque, and Ahmed]{hassan2021comparative}
Kazi Md~Abir Hassan, Md~Atiqul Haque, and Sakif Ahmed.
\newblock Comparative study of forecasting global mean sea level rising using
  machine learning.
\newblock In \emph{2021 International Conference on Electronics, Communications
  and Information Technology (ICECIT)}, pages 1--4. IEEE, 2021.

\bibitem[Kay et~al.(2015)Kay, Deser, Phillips, Mai, Hannay, Strand, Arblaster,
  Bates, Danabasoglu, Edwards, et~al.]{kay2015community}
Jennifer~E Kay, Clara Deser, A~Phillips, A~Mai, Cecile Hannay, Gary Strand,
  Julie~Michelle Arblaster, SC~Bates, Gokhan Danabasoglu, James Edwards, et~al.
\newblock The community earth system model (cesm) large ensemble project: A
  community resource for studying climate change in the presence of internal
  climate variability.
\newblock \emph{Bulletin of the American Meteorological Society}, 96\penalty0
  (8):\penalty0 1333--1349, 2015.

\bibitem[Lundberg and Lee(2017)]{lundberg2017unified}
Scott~M Lundberg and Su-In Lee.
\newblock A unified approach to interpreting model predictions.
\newblock \emph{Advances in neural information processing systems}, 30, 2017.

\bibitem[Maher et~al.(2019)Maher, Milinski, Suarez-Gutierrez, Botzet, Dobrynin,
  Kornblueh, Kr{\"o}ger, Takano, Ghosh, Hedemann, et~al.]{maher2019max}
Nicola Maher, Sebastian Milinski, Laura Suarez-Gutierrez, Michael Botzet,
  Mikhail Dobrynin, Luis Kornblueh, J{\"u}rgen Kr{\"o}ger, Yohei Takano, Rohit
  Ghosh, Christopher Hedemann, et~al.
\newblock The max planck institute grand ensemble: enabling the exploration of
  climate system variability.
\newblock \emph{Journal of Advances in Modeling Earth Systems}, 11\penalty0
  (7):\penalty0 2050--2069, 2019.

\bibitem[M{\"u}ller et~al.(2018)M{\"u}ller, Jungclaus, Mauritsen, Baehr,
  Bittner, Budich, Bunzel, Esch, Ghosh, Haak, et~al.]{muller2018higher}
Wolfgang~A M{\"u}ller, Johann~H Jungclaus, Thorsten Mauritsen, Johanna Baehr,
  Matthias Bittner, R~Budich, Felix Bunzel, Monika Esch, Rohit Ghosh, Helmut
  Haak, et~al.
\newblock A higher-resolution version of the max planck institute earth system
  model (mpi-esm1. 2-hr).
\newblock \emph{Journal of Advances in Modeling Earth Systems}, 10\penalty0
  (7):\penalty0 1383--1413, 2018.

\bibitem[Nerem et~al.(2018)Nerem, Beckley, Fasullo, Hamlington, Masters, and
  Mitchum]{nerem2018climate}
Robert~S Nerem, Brian~D Beckley, John~T Fasullo, Benjamin~D Hamlington, Dallas
  Masters, and Gary~T Mitchum.
\newblock Climate-change--driven accelerated sea-level rise detected in the
  altimeter era.
\newblock \emph{Proceedings of the national academy of sciences}, 115\penalty0
  (9):\penalty0 2022--2025, 2018.

\bibitem[Nieves et~al.(2021)Nieves, Radin, and
  Camps-Valls]{nieves2021predicting}
Veronica Nieves, Cristina Radin, and Gustau Camps-Valls.
\newblock Predicting regional coastal sea level changes with machine learning.
\newblock \emph{Scientific Reports}, 11\penalty0 (1):\penalty0 1--6, 2021.

\bibitem[Sinha et~al.(2022)Sinha, Monteleoni, Fasullo, and Nerem]{sinha2022sea}
Saumya Sinha, Claire Monteleoni, John Fasullo, and R~Steven Nerem.
\newblock Sea-level projections via spatiotemporal deep learning from altimetry
  and cesm large ensembles.
\newblock In \emph{Fall Meeting 2022}. AGU, 2022.

\bibitem[Sun et~al.(2020)Sun, Wan, and Liu]{sun2020estimation}
Qinting Sun, Jianhua Wan, and Shanwei Liu.
\newblock Estimation of sea level variability in the china sea and its vicinity
  using the sarima and lstm models.
\newblock \emph{IEEE Journal of Selected Topics in Applied Earth Observations
  and Remote Sensing}, 13:\penalty0 3317--3326, 2020.

\end{thebibliography}

\end{document}